\documentstyle[12pt,epsfig,amssymb]{article}

\def\gapprox { \lower.7ex \hbox {$\;\stackrel{\textstyle >}{\sim}\;$}}

\newcommand{\chupiii}{$\chi U\phi_3$}
\newcommand{\cbcex}{\langle\overline{\chi}\chi\rangle}

\newlength{\capsep}
\newcommand\fdfig[1]{%
  \psfig{file=#1,angle=90,width=.9\hsize,bbllx=65,bblly=15,bburx=540,bbury=780}}%

\newcommand\fdiifig[1]{%
  \psfig{file=#1,angle=90,width=.9\hsize,bbllx=65,bblly=25,bburx=540,bbury=810}}%

\begin{document}
\vspace{5mm}
\begin{center}
  {\bf ~ \hfill HLRZ1998\_11\\ 
    Strongly coupled lattice gauge theory\\ with
    dynamical fermion mass generation\\ in three dimensions}
\end{center}
\vspace{5mm}
\begin{center}
I. M. Barbour$^*)$ and N. Psycharis$^*)$\\
Dept. of Physics and Astronomy\\
University of Glasgow, Glasgow G12 8QQ, U.K.\\
\vspace{4mm}
and\\
\vspace{4mm}
E. Focht, W. Franzki and J. Jers{\'a}k\\
Inst.\ f. Theor.\ Physik E\\
RWTH Aachen, D-52056 Aachen, Germany
\end{center}
\vspace{25mm}
\begin{abstract}
  
  We investigate the critical behaviour of a three-dimensional lattice
  \chupiii\ model in the chiral limit. The model consists of a staggered
  fermion field, a $U(1)$ gauge field (with coupling parameter $\beta$) and a
  complex scalar field (with hopping parameter $\kappa$).  Two different
  methods are used: 1) fits of the chiral condensate and the mass of the
  neutral unconfined composite fermion to an equation of state and 2) finite
  size scaling investigations of the Lee-Yang zeros of the partition function
  in the complex fermion mass plane. For strong gauge coupling ($\beta < 1$)
  the critical exponents for the chiral phase transition are determined.  We
  find strong indications that the chiral phase transition is in one
  universality class in this $\beta$ interval: that of the three-dimensional
  Gross-Neveu model with two fermions. Thus the continuum limit of the
  \chupiii\ model defines here a nonperturbatively renormalizable gauge theory
  with dynamical mass generation. At weak gauge coupling and small $\kappa$,
  we explore a region in which the mass in the neutral fermion channel is
  large but the chiral condensate on finite lattices very small. If it does
  not vanish in the infinite volume limit, then a continuum limit with massive
  unconfined fermion might be possible in this region, too.

\vspace{0.3cm}  
\end{abstract}

$^*)$ UKQCD Collaboration
\newpage
\section{Introduction}
\label{sec:intro}

Strongly coupled gauge theories are interesting candidates for new
mass generating mechanisms because they tend to break chiral symmetry
dynamically.  However, the fermions which acquire mass through this
mechanism usually get confined. It was pointed out \cite{FrJe95a} that
this can be avoided in a class of chiral symmetric strongly coupled
gauge theories on the lattice in which the gauge charge of the fermion
is shielded by a scalar field of the same charge. The question is,
whether these models are nonperturbatively renormalizable at strong
gauge coupling such that the lattice cutoff can be removed. If so, the
resulting theory might be applicable in continuum and constitute a
possible alternative to the Higgs mechanism \cite{FrJe95a}.

In this work we investigate such a lattice model in three dimensions
with a vectorlike U(1) gauge symmetry, which we call \chupiii\ model.
It consists of a staggered fermion field $\chi$ with a global U(1)
chiral symmetry, a gauge field $U\in\mbox{U(1)}$ living on the lattice
links of length $a$ and a complex scalar field $\phi$ with frozen
length $|\phi|=1$. It is characterized by the dimensionless gauge
coupling parameter $\beta$ (proportional to the inverse squared
coupling constant), the hopping parameter $\kappa$ of the scalar field
and the bare fermion mass $am_0$. The unconfined fermion is the
composite state $F=\phi^\dagger\chi$.  In a phase with broken chiral
symmetry, it has nonvanishing mass $am_F$ in the chiral limit $m_0=0$.
The \chupiii\ model can be seen either as a generalization of
three-dimensional compact QED with a charged scalar field added or as
three-dimensional U(1) Higgs model with added fermions.

The same model has also been investigated in two and four dimensions.  In two
dimensions it seems to be in the universality class of the Gross-Neveu model
\cite{FrJe96c} at least for strong gauge coupling, thus being renormalizable.
Therefore the shielded gauge-charge mechanism of dynamical mass generation
suggested in \cite{FrJe95a} works in two dimensions and its long range
behaviour is equivalent to the four fermion theory. In four dimensions there
is also a region in $\beta$ ($0\leq\beta<0.64$) in which the model behaves in
a very similar manner to the corresponding four-fermion theory, the
Nambu--Jona-Lasinio model with a massive fermion whose mass scales at the
critical point \cite{FrFr95}. Here both models belong to the same universality
class and have the same renormalizability properties. But for intermediate
coupling there evidently exists a special point. It is a tricritical point at
which, together with the composite fermion $F$, scaling of a particular scalar
state was found.  This composite scalar can be interpreted as a gauge ball
mixing with a $\phi^\dagger$-$\phi$ state. Thus the gauge degrees of freedom
play an important dynamical role and the model belongs to a new universality
class of models with dynamical mass generation, whose renormalizability is of
much interest \cite{FrJe98a,FrJe98b}.

\begin{figure}
  \begin{center}
    \leavevmode
    \fdfig{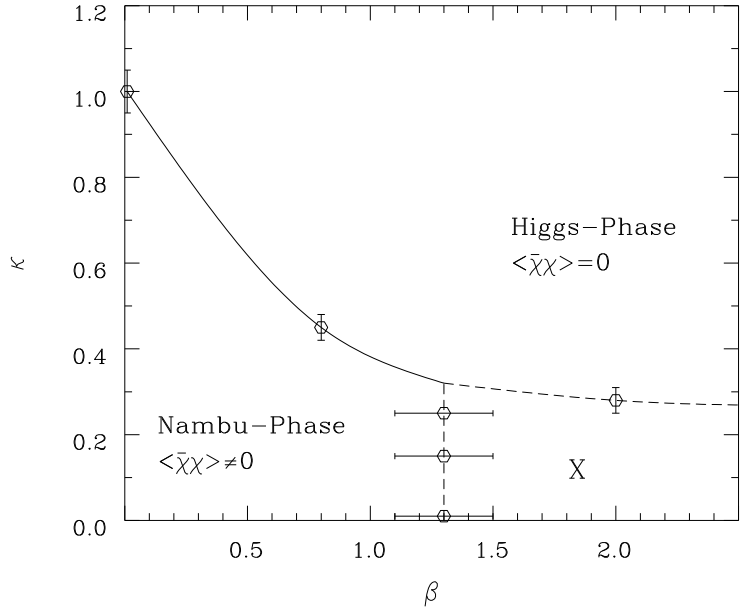}\hspace*{-2cm}%
    \vspace{\capsep}%
    \caption[xxx]{%
      Phase diagram of the \chupiii\ model for $m_0=0$. For $\beta < 1$ a
      clear phase transition between the Nambu and Higgs phases can be
      observed.  Whether the region X at large $\beta$ and small $\kappa$
      forms a third phase or belongs to one of the other phases, separated
      from it only by a crossover, is discussed in the text. All phase
      transitions seem to be 2$^{\rm nd}$ order.}
    \label{fig:pd}
  \end{center}
\end{figure}

In this paper we investigate the phase diagram and the critical
behaviour of the model in three dimensions. We find that in the chiral
limit $m_0 = 0$ the \chupiii\ model has three regions in the $\beta-\kappa$
plane with different properties with respect to the chiral symmetry.
They are indicated in Fig.~\ref{fig:pd}.  The region at strong gauge
coupling (small $\beta$) and small $\kappa$ is the Nambu phase where
the chiral symmetry is broken and the neutral fermion $F$ is massive.
At large $\kappa$ chiral symmetry is restored and the fermion $F$ is
massless. This phase is labelled the Higgs phase because of its
properties in the weak coupling limit. The third is the X region at
large $\beta$ and small $\kappa$. It is conceivable that this region
is analytically connected with either the Nambu or Higgs phase but it
may well be a separate phase. In this region the mass measured in the
fermion channel is large, but the chiral condensate is very small
(within our numerical accuracy consistent with zero).

The main result of our paper is the determination of the critical
behaviour at strong gauge coupling. We find strong indications that
the chiral phase transition between the Nambu and Higgs phases is in
one universality class for all $\beta\lesssim 1$. It is the class of
the three-dimensional Gross-Neveu model which is known to be
(nonpertubatively) renormalizable \cite{RoWa89b}.  That model is the
$\beta=0$ limit of the \chupiii\ model \cite{LeShr87a}.  This
universality means that the continuum limit of the \chupiii\ model
defines a nonperturbatively renormalisable gauge theory in which the
fermion mass is generated dynamically by the shielded gauge-charge
mechanism. However, it also means that in this $\beta$ region the
gauge field is auxiliary and the \chupiii\ model does not
represent a new class of field theories.

The chiral properties of the region X are elusive and their
determination would require substantial effort and resources. This is
beyond the scope of the present work and we made only an exploratory
investigation. But we point out that, provided the chiral symmetry is
broken there, the phase transition between the region X and the
Higgs phase gives rise to another possible construction for a
continuum theory containing an unconfined fermion with dynamically
generated mass. It could continue to be in the universality class of
the three-dimensional Gross-Neveu model. But experience
\cite{FrJe98a,FrJe98b} with the four-dimensional model in the vicinity
of the tricritical point suggests that at larger $\beta$ the gauge
degrees of freedom are dynamical and a new universality class may be
present. This interesting possibility, and the possible existence of a
tricritical point in three dimensions, deserves further study.

Our investigation is mainly based on two methods: firstly, via fits to
an equation of state and, secondly, via a finite size scaling
investigation of the Lee-Yang zeros in the complex fermion mass plane.
The investigation of a phase transition via fits to an equation of
state is quite reliable because the finite size effects we find close
to the phase transition are usually small. Therefore we expect a
simple finite size scaling, describe it by an empirical formula and
extrapolate observables to the infinite volume. Then we do a
simultaneous fit to the fermion mass $am_F$ and the chiral condensate
$\cbcex$.

As first pointed out by Lee and Yang \cite{YaLe52,LeYa52}, the determination
of the finite size scaling behaviour of the complex zeros of a partition
function could be a direct method for the determination of the critical
properties of the associated theory. In this paper we investigate these zeros
of the canonical partition function in the complex bare fermion mass plane.
These zeros control the fermion condensate and its associated susceptibilities
\cite{BaBe93b,BaBu95}, physical quantities which are often measured directly
on the lattice and used, via finite size scaling, to determine the critical
behaviour.

In the region X, where the chiral condensate is very small, both
methods fail to provide reliable results. A small condensate suggests
that the Lee-Yang zeros cannot be near to the physical region.
Nevertheless, it is of interest to investigate if the closest zeros
can be determined with sufficient accuracy to ascertain their finite
size scaling (and hence that of the condensate).

The paper is organized as follows. In the next section we introduce the model
in detail, define the observables we use and briefly summarize the method of
the Lee-Yang zeros. In section~\ref{sec:strong} we present evidence for the
universality at strong gauge coupling. In section~\ref{sec:weak} we present
the results obtained at weak coupling and discuss their possible
interpretations. In the last section our results are summarized.

\section{The model}

The \chupiii\ model is defined on a 3-dimensional cubic lattice with
periodic boundary conditions except for antiperiodic boundary
conditions for the fermion field in the ``time'' direction. The action
reads:
\begin{equation}
  \label{chupact}
  S_{\chi U \phi} = S_\chi + S_U + S_\phi,
\end{equation}
with
\begin{eqnarray*}
  S_\chi &=& \frac{1}{2} \sum_x \overline{\chi}_x
  \sum_{\mu=1}^3 \eta_{\mu x} (U_{x,\mu} \chi_{x+\mu} - U^\dagger_{x-\mu,\mu}
  \chi_{x-\mu})
  +am_0 \sum_x \overline{\chi}_x \chi_x \;,\\ 
  S_U &=& \beta \sum_{x,\mu<\nu} (1-{\rm Re}\,{U_{x,\mu\nu}}) \;,\\ 
  S_\phi &=& - \kappa \sum_x \sum_{\mu=1}^3
  (\phi^\dagger_x U_{x,\mu} \phi_{x+\mu} + {\rm h.\,c.}) \;.
\end{eqnarray*}

Here $\chi_x$ are the Kogut-Susskind fermion fields with $\eta_{\mu
  x}=(-1)^{x_1+\cdots+x_{\mu-1}}$. Because of doubling our model describes two
four-component fermions ($N_f=2$). The bare mass $am_0$ of the fermion is
introduced for technical reasons. We are interested in the chiral limit
$m_0=0$.  The $a$ in front of $m_0$ indicates that we have to distinguish
between the chiral limit in the continuum ($m_0=0$) and the continuum limit of
the lattice model, where $am_0\rightarrow 0$ can also be achieved by
$a\rightarrow 0$ at nonzero $m_0$.

$U_{x,\mu}$ represents the compact link variable and $U_{x,\mu\nu}$ is
the plaquette product of the link variables $U_{x,\mu}$.   

The hopping parameter $\kappa$ vanishes, if the square of the bare
mass of the scalar field is $+\infty$, and is infinite if the bare
mass squared is $-\infty$. The scalar field $\phi$ has frozen length
$|\phi|=1$. This choice is made in order to restrict the number of
parameters of the model. Without that, symmetries and dimensionality
of couplings would allow several other terms in the action.

We stress that the charges of the fundamental fields exclude a direct Yukawa
coupling between the fundamental fields.

The model has some interesting limiting cases. For $\kappa=0$ the
scalar field decouples and the model is equivalent to
three-dimensional compact QED with fermions. It is known
\cite{Po75,Po77,BaMy77,GoMa82} that pure compact QED has no phase
transition and, as $\beta \rightarrow \infty$, it is confining via a
linear potential with an exponentially decreasing string tension.
There is an indication that, with fermions, chiral symmetry is broken
at large coupling, but at weak coupling results are inconclusive
\cite{BuIr88}. It has been suggested that, in noncompact QED with
fermions, the phase diagram is dependent on the number of flavors and
that, at weak coupling, chiral symmetry is broken only for a small
number (less than about 3.5) of fermions
\cite{ApNa88,DaKo89,DaKo90,HaKo90}. (A recent description of the
status of three-dimensional QED can be found in \cite{GuHa96,HaOi98}.)
It is quite probable that, at weak coupling, both the compact and
non-compact formulations have quite similar properties. If so, then
these (uncertain) features suggest that the chiral symmetry is broken
in the $\kappa = 0$ limit of the phase X and thus presumably in the
whole phase X.

In the weak gauge coupling limit, $\beta=\infty$, the fermions are
free with mass $am_0$, and $S_\phi$ reduces to the XY$_3$ model. It
has a phase transition at $\kappa\approx 0.27$.

At $am_0 = \infty$ the model reduces to the three-dimensional compact U(1)
Higgs model. For its recent investigation with numerous references see
\cite{KaKa98}.

For $\beta=0$ the gauge and scalar fields can be integrated out exactly
\cite{LeShr87a} and one ends up with a lattice version of the
three-dimensional four-fermion model  
\begin{equation}
     S_{4f} = -\sum_x \sum_{\mu=1}^3
      [ G \overline{\chi}_x \chi_x \overline{\chi}_{x+\mu} \chi_{x+\mu}
      -\frac{1}{2} \eta_{\mu x} (\overline{\chi}_x \chi_{x+\mu}
                                 - \overline{\chi}_{x+\mu} \chi_x) ]
      +\frac{am_0}{r} \sum_x \overline{\chi}_x \chi_x,
\label{S-4-fermion}
\end{equation} 
the parameters $G$ and $r$ being related to $\kappa$ \cite{LeShr87a,FrJe95a}.

We refer to this model as the Gross-Neveu model.  Some caution is in
place, however. There is some uncertainty whether the four-fermion
model (\ref{S-4-fermion}) is a lattice version of the Gross-Neveu
model or of the Thirring model. The four-fermion action
(\ref{S-4-fermion}) was used in four dimensions for the study of the
Nambu--Jona-Lasinio model e.g. in \cite{BoKe89,AlGo95}, which would
correspond to the Gross-Neveu model in three dimensions. Recently in
\cite{DeHa96,DeHa97} the four-fermion action (\ref{S-4-fermion}) in
three dimensions is interpreted as the Thirring model and similar
interpretation is implied by K.-I. Kondo \cite{Ko95}. For our number
of fermions, $N_f=2$, the distinction might be unimportant and both
models might actually coincide\footnote{J.J. thanks M.  G\"ockeler,
  S.J. Hands, and K.-I. Kondo for discussions on these questions. Some of
  them are exposed in \cite{DeHa97} }. The Gross-Neveu model has a
chiral phase transition and is nonperturbatively renormalisable (see
\cite{RoWa89b,RoWa91} and references therein). The properties of the
$N_f=2$ Thirring model appear to be similar \cite{Ko95,DeHa96,DeHa97}.
For our purposes the important property of the three-dimensional
four-fermion model obtained in the $\beta = 0$ limit of the \chupiii\ 
model is its renormalisability, which presumably holds for both
interpretations.

\subsection{Observables}

Because we are interested in the chiral properties of the model we concentrate
on the chiral condensate and the fermion mass.

The chiral condensate is defined by
\begin{equation}
  \cbcex = \left\langle {\rm Tr}\, M^{-1}\right\rangle
\end{equation}
where $M$ is the fermion matrix. The trace is measured with a gaussian
estimator.

The physical fermion of the \chupiii\ model is the gauge invariant
composite fermion $F=\phi^\dagger\chi$. We measure its mass $am_F$ in
momentum space with the usual procedure, as described (for the model
in 4 dimensions) in \cite{FrFr95}. We checked that the results are in
good agreement with the fits done in configuration space.
In the three-dimensional model we find the fits to
$G^{1\!\!1}$ to be the most stable, so we use them for the data shown in
this paper.

Both observables need to be extrapolated to infinite volume. This
procedure is described in section
\ref{sec:infvol}.

We remark that the required numerical effort for the study of the
\chupiii\ model was very high. We needed significantly more matrix
inversions than for the four-dimensional case \cite{FrJe98b} at the
same volume and $am_0$. Their number also depends significantly on
$\beta$: The simulation at $\beta=0$ required about O(1000)
conjugate-gradient steps, about 2-5 times more than at larger
$\beta$-values.  Surprisingly, the number of required steps scattered
in a very broad interval. The maximal step-number was at least a
factor of 2-3 above the average. This might be connected with the
observation, that the chiral condensate has a very asymmetric
distribution.

\subsection{Equation of State}
\label{sec:eosdef}
A standard way to analyse the critical exponents of a chiral phase transition
is via the use of an equation of state (EOS). Normally data close to the phase
transition can be well described by such an ansatz. In our model for fixed
$\beta$ this equation reads
\begin{equation}
  \label{eq:eoscbc}
  am_0 = \cbcex^\delta F\left( (\kappa-\kappa_c) \cbcex^{-1/\beta_\chi}
  \right)\;,\quad F(x)=Rx+S\,.
\end{equation}
Here $\cbcex$ is the infinite volume value of the chiral condensate for given
$am_0$, $\kappa$ and $\beta$.  $\beta_\chi$ and $\delta$ are the exponents
defined in analogy to a magnetic transition. The index $\chi$ is added to
distinguish the exponent and the coupling.  The scaling function $F$ is used in
its linear approximation and $R$ and $S$ are free constants. We apply this
equation in the region for which $\kappa\approx\kappa_c$ and where we might
expect the scaling deviations to be small.

\begin{figure}
  \begin{center}
    \psfig{file=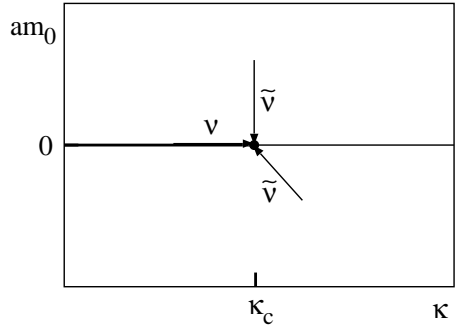,width=6cm}
    \caption{First order chiral phase transition line (bold line) and the
      critical point in a plane $\beta=const$. The fermion mass $am_F$ scales
      with exponent $\nu$ in the direction tangential to the
      transition line and with exponent $\tilde\nu$ in any other direction.}
    \label{fig:chipt}
  \end{center}
\end{figure}

It is also useful to assume the corresponding scaling equation 
for the fermion mass:
\begin{equation}
  \label{eq:eosmf}
  am_0 = (am_F)^{1/\tilde\nu} G\left( (\kappa-\kappa_c) (am_F)^{-1/\nu}
  \right)\;,\quad G(x)=Ax+B\,.
\end{equation}
The exponent $\nu$ is the correlation length critical exponent in the
chiral plane ($am_0=0$).  $\tilde\nu$ is an analogous exponent
obtained if one approaches the critical point from outside the chiral
plane. The two exponents have to be distinguished. At fixed $\beta$ in
the chiral plane ($am_0=0$) the fermion mass scales with $\nu$:
$am_F\propto (\kappa-\kappa_c)^{\nu}|_{am_0=0}$, whereas for all other
straight paths into the critical point (for example $\kappa=\kappa_c$)
it scales with $am_0$ as: $am_F \propto
{am_0}^{\tilde\nu}|_{\kappa-\kappa_c\propto am_0}$.  This is indicated
in figure~\ref{fig:chipt} in the plane $\beta=const$.

Figure~\ref{fig:chipt} also illustrates that for $\kappa<\kappa_c$ the
chiral condensate changes sign and makes a jump if one crosses the
line $am_0=0$.  This means that it is a line of first order phase
transitions.  For $\kappa>\kappa_c$ the line becomes a line of second
order phase transitions on which the fermion mass gets critical. In
between there is a critical point ($\kappa=\kappa_c$).

If hyperscaling holds, only two of the four exponents defined by the
equations of state are independent. The corresponding scaling
relations are:
\begin{equation}
  \label{eq:scal}
  \delta = \frac{1}{d\tilde\nu - 1} \quad\mbox{and}\quad
  \beta_\chi=\nu\left(d-\frac{1}{\tilde\nu}\right)\,,
\end{equation}
where $d=3$ is the space-time dimension.

\subsection{Lee-Yang zeros}
\label{sec:deflyz}
The canonical partition function, after integration over the Grassmann
variables and using the irrelevance of overall multiplicative factors,
can be defined as:
\begin{equation}
  Z(\beta,\kappa,am_{0}) = \frac{\int dU d\phi \det 
    M[am_{0},U] e^{S_{U\phi}}}{\int dU d\phi \det
    M[a\hat m_0,U] e^{S_{U\phi}}}.
\end{equation}
Here $S_{U\phi} = S_U+ S_\phi$, $M$ is the usual fermionic matrix
for Kogut-Susskind fermions and $a\hat m_0$ is some (arbitrary)
``updating'' fermion mass at which the ensemble of gauge fields is
generated.

Since the mass dependence of $M$ is purely diagonal, the partition
function can be written as the average over the ensemble of the
characteristic polynomials of $M$, i.\,e.:
\begin{eqnarray}
  Z(\beta,\kappa,am_{0}) &=& \left\langle \frac{ \sum_{n=0}^{V \over 2}
    C_n[U[\beta,\kappa]] (am_{0})^{2n}}
    { \det M[a\hat m_0,U[\beta,\kappa]]}\right\rangle_{a\hat m_0}\\
  \label{eq:polynom}
  &=& \sum_{n=0}^{V \over 2} A_n[\beta,\kappa] (am_{0})^{2n}.
\end{eqnarray}

The coefficients $C_n$ of the characteristic polynomial are obtained
from the eigenvalues of $M[0,U]$ which are imaginary and appear in
complex conjugate pairs. In the simulations described below they were
obtained using the Lanczos algorithm.

The Lee-Yang zeros are the zeros of this polynomial representation of
the partition function.  The zeros were found by using a standard root
finding algorithm on the equivalent sets of polynomials generated as
in Eq.~($8$):
\begin{equation}
  \sum_{n=0}^{V \over 2} A_n^i (am_{0}^2-a\bar m_i^2)^n
\end{equation}
for a set of $a\bar m_i$ in the region where we expect the lowest
zeros to occur.  This allowed us to avoid the problems associated with
rounding errors in the root-finder.  We required that a zero be found
consistently for the subset of the $a\bar m_i$ closest to it. These
zeros in the bare mass we label as $y_i$ in the following.

The errors in the Lee-Yang zeros are estimated by a Jacknife method.
The coefficents for each lattice size were averaged to produce $6$
subsets of averaged coefficients, each taking into account $5/6$ of
the measurements.  These $6$ different sets of coefficients give $6$
different results for the Lee-Yang zeros from which the variance was
calculated.

The critical properties of the system are determined by the zeros
lying closest to the real axis.  The zero with the smallest imaginary
part we label $y_1$. It is also called edge singularity.  With
increasing finite volume it converges to the critical point.  For a
continuous phase transition the position of the zeros closest to the
real axis in the complex plane is ruled by the scaling law
\begin{equation}
  y_{i}(\beta,\kappa,L) - y_{R}(\beta,\kappa,\infty) = A_i L^{-1/s},
  \label{eq:scaly}
\end{equation}
where the $A_i$'s are complex numbers.  The exponent
$s=s(\beta,\kappa)$ describes the finite size scaling of the
correlation length.  For our model $y_{R}(\beta,\kappa,\infty)=0$ and
we ignore it in the following.

It immediately follows that the real and the imaginary parts of the
zeros should scale independently with the same exponent.  In
particular, for the zero $y_1$ closest to the chiral phase transition
(at $am_0=0$)
\begin{equation}
  \label{eq:scalim}
  {\rm Im}\, y_{1}(\beta,\kappa,L) = A_{I} L^{-1/s},
\end{equation}
with a similar scaling behaviour for ${\rm Re}\,y_{1}(\beta,\kappa,L)$
via $A_{R}$.  In practice the real part of the zero is much smaller
than its imaginary part or is identically zero. So
eq.~(\ref{eq:scalim}) usually provides a more reliable measure of the
exponent than the scaling of the real part.

Although the above scaling law was originally established for the case
of a continuous phase transition, it can also be extended to that of a
first order phase transition.  Since there is no divergent correlation
length, the exponent is determined only by the actual dimension of the
system.  In this case, for a three-dimensional model we expect
$s=\frac{1}{3}$.

\begin{figure}
  \begin{center}
    \psfig{file=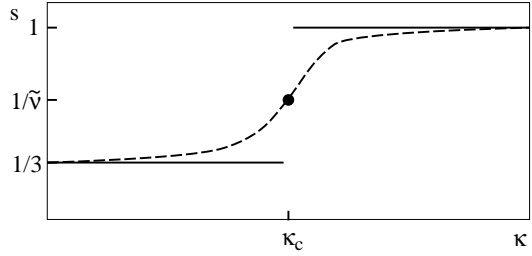,width=7cm}
    \caption{Schematic plot of the exponent $s$ in infinite volume (full lines
      and dot) and the effective $s$ in finite volume (dashed line).}
    \label{fig:effnu}
  \end{center}
\end{figure}

At the critical point ($\kappa=\kappa_c$) we expect $s$ to be equal to
$\tilde\nu$, because the fermion correlation length should be the
relevant one. In the symmetric phase ($\kappa>\kappa_c$) we expect
scaling with $s=1$, because $am_F\propto am_0$.  This behaviour is
indicated in Fig.~\ref{fig:effnu} by the full lines and the dot.

\begin{figure}
  \begin{center}
    \psfig{file=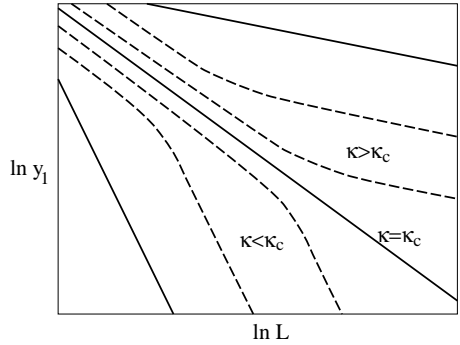,width=6cm}
    \caption{Expected finite size scaling of the zero $y_1$ with lattice size.}
    \label{fig:fsscal}
  \end{center}
\end{figure}%
In practice it is important to understand the scaling deviations on a
finite lattice. The expected behaviour is shown schematically in
Fig.~\ref{fig:fsscal}. Far away from the critical point we expect
linear scaling in the log-log plot with $s=1/3$ in the broken phase,
and $s=1$ in the symmetric phase.  At the critical point we expect
linear scaling and the exponent should be $s=\tilde\nu$. These
expectations are indicated by the full lines.  Close to the phase
transition, we expect a crossover. For small lattice sizes the
exponent should be close to $\tilde\nu$ and then change to $1/3$ and
$1$, respectively, if the lattice size is increased and the true
scaling shows up.  This is indicated in Fig.~\ref{fig:fsscal} by the
dashed lines. For a set of lattice sizes this defines an effective $s$
which smoothly goes through $\tilde\nu$ at the critical point. Such an
effective $s$ is represented in Fig.~\ref{fig:effnu} by a dashed line.

Therefore, in order that the critical exponent can be determined, we must either
know the position of the critical point accurately or have many simulations on
large lattices so that the scaling deviations can be measured accurately.  In
practice the limited knowledge of the position of the critical point leads to
the largest uncertainty in the determination of $\tilde\nu$ by this method.

\section{Universality at Strong Coupling}
\label{sec:strong}
At strong coupling the chiral phase transition can be seen clearly and we
investigate the scaling behaviour and the universality along this line.

We determined the Lee-Yang zeros, the chiral condensate and the fermion mass
for various values of $\kappa$ at $\beta=0.00$ and 0.80.  In this section we
want to investigate how the transition changes as $\beta$ increases from zero.
Therefore we have investigated the scaling of the data at $\beta=0$, i.e. the
four-fermion model, as a reference and compare it with the scaling found
at $\beta=0.80$.

\subsection{Equation of State}

\label{sec:infvol}
\begin{figure}
  \begin{center}
    \fdfig{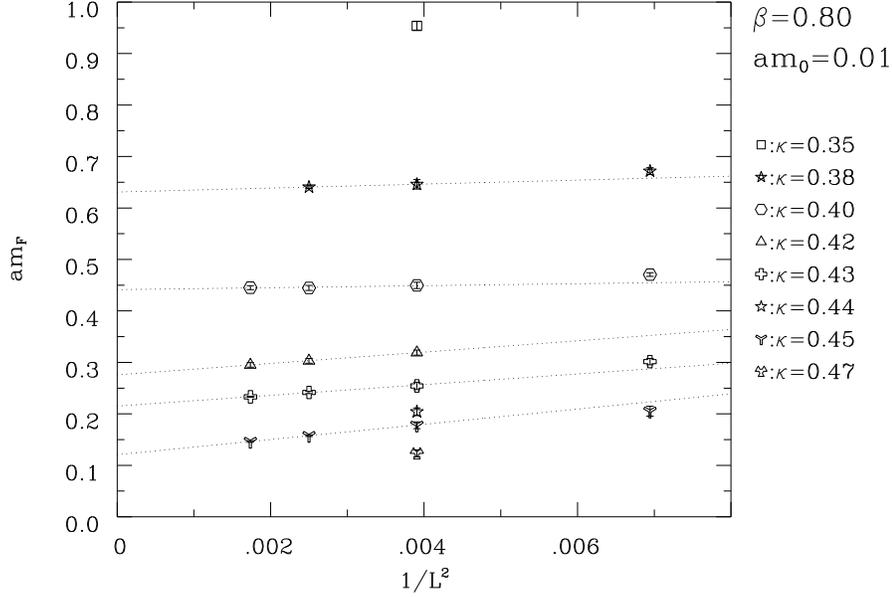}
    \vspace{\capsep}%
    \caption[xxx]{%
       Data for $am_F$ at $\beta=0.80$ and $am_0=0.01$ plotted against
      $1/L^2$. The dotted lines are a fit with eq.~(\protect\ref{eq:2L}) to
      the data with $L\geq 16$.}
    \label{fig:amext}
  \end{center}
\end{figure}

Here we determine the critical exponents of the chiral phase transition by
using the EOS for $am_F$ and $\cbcex$. Although we did simulations on lattices
up to $24^3$, our conclusions still depend to some extent on our choice of
ansatz for the extrapolation of $am_F$ and $\cbcex$ to infinite volume.
Fig.~\ref{fig:amext} shows, as an example, our data for $am_F$ at $\beta=0.80$
and $am_0=0.01$ plotted against $1/L^2$.

For the extrapolation we tried three approaches:
\begin{eqnarray}
  \label{eq:2L}
  am_F(L) &=& am_F(\infty) + A \frac{1}{L^2}\,,\\
  \label{eq:1L}
  am_F(L) &=& am_F(\infty) + A \frac{1}{L}\,,\\
  \label{eq:1Lexp}
  am_F(L) &=& am_F(\infty) + A \frac{1}{L}\exp(-am_F(\infty)\,L)\,.
\end{eqnarray}
Each has two free parameters: $am_F(\infty)$ and $A$.  To judge the
quality of the fits we first compared the $\chi^2$ per degree of
freedom using our data on $16^3$, $20^3$ and $24^3$ lattices. This was
done at the values of $\beta$ and $\kappa$ at which we have good
statistics. Our results are shown in table~\ref{tab:amext}. It turned
out that the results for these lattice sizes are not conclusive as to
which extrapolation formula should be used, because, for each ansatz,
all $\chi^2$ per degree of freedom are usually below 1.
\begin{table}
  \begin{center}
    \begin{tabular}{|lll|ll|ll|ll|} \hline
      &&& \multicolumn{2}{c|}{Fit 1} &
      \multicolumn{2}{c|}{Fit 2} & \multicolumn{2}{c|}{Fit 3} \\
      $\beta$ & $\kappa$ & $am_0$ & $am_F$ & $\chi^2$ & $am_F$ & $\chi^2$ &
      $am_F$ & $\chi^2$\\ \hline 
      0.00&0.95&0.01 & 0.269(3) & 0.30 & 0.256(7)  & 0.53 & 0.278(2) & 0.01 \\
      0.00&1.00&0.01 & 0.194(4) & 0.22 & 0.180(9)  & 0.45 & 0.202(2) & 0.01 \\
      0.00&1.05&0.01 & 0.132(5) & 0.23 & 0.111(10) & 0.04 & 0.142(3) & 0.62 \\
      0.00&0.95&0.02 & 0.337(3) & 0.41 & 0.327(6)  & 0.22 & 0.343(1) & 2.07 \\
      \hline
      0.80&0.40&0.01 & 0.441(8) & 0.12 & 0.437(2)  & 0.15 & 0.445(3) & 0.03 \\
      0.80&0.42&0.01 & 0.276(8) & 0.01 & 0.247(16) & 0.08 & 0.295(4) & 0.31 \\
      0.80&0.43&0.01 & 0.215(5) & 0.07 & 0.191(9)  & 0.01 & 0.230(2) & 1.33 \\
      0.80&0.45&0.01 & 0.122(2) & 0.01 & 0.087(1)  & 0.14 & 0.137(3) & 0.36 \\
      \hline
    \end{tabular}
    \caption{%
      Results of the fits to the finite size behaviour at different couplings
      and masses on $16^3$, $20^3$, $24^3$ lattices.  The extrapolated
      infinite volume mass $am_F=am_F(\infty)$ and the $\chi^2$ per degree of
      freedom for the three fits are given: with eq.~(\protect\ref{eq:2L})
      (Fit 1), eq.~(\protect\ref{eq:1L}) (Fit 2) and
      eq.~(\protect\ref{eq:1Lexp}) (Fit 3).}
    \label{tab:amext}
  \end{center}
\end{table}

However, the fit with eq.~(\ref{eq:2L}) is significantly prefered if
compared with the $12^3$ data. We therefore adopted this fit for our
extrapolations, but data from the $12^3$ lattice was not included.
Such an extrapolation is indicated in Fig.~\ref{fig:amext} by the
dotted lines.

For the chiral condensate the finite size effects are in general
smaller and with opposite sign. Again, consideration of the $12^3$
lattices favoured a fit ansatz analogous to eq.~(\ref{eq:2L}).

We describe in detail the analysis in which we used the ansatz of
eq.~(\ref{eq:2L}) to extrapolate all our data for $am_F$ and $\cbcex$,
obtained on $16^3$ and larger lattices, to infinite volume. The error
was calculated with the MINOS routine from the MINUIT library.  All
results presented in the following change somewhat quantitatively, but
not qualitatively, if a different extrapolation formula is used.

The data at different $\kappa$ and $am_0$, extrapolated to the
infinite volume, were analyzed by means of the EOS. We included only
the data at $am_0=0.01$ and $am_0=0.02$. The chosen $\kappa$ range was
0.80\,\ldots\,1.05 for $\beta=0.00$ and 0.38\,\ldots\,0.47 for
$\beta=0.80$.

As a first step we analyzed the data for $am_F$ and $\cbcex$
independently and fitted to their corresponding EOS (\ref{eq:eosmf})
and (\ref{eq:eoscbc}). The results are given in table
\ref{tab:eossep}. As can be seen, for both $\beta$'s the critical
$\kappa$ values $\kappa_c$ are identical within the error bars.
\begin{table}
  \begin{center}
    \renewcommand{\arraystretch}{1.2}
    \begin{tabular}{c|l|llllll|}
      \cline{2-8}
      &$\beta$ & $\kappa_c$ & $\nu$ & $\tilde\nu$ & $A$ & $B$ & $\chi^2$ \\
      \cline{2-8} $am_F$:
      &0.00 & 0.987(33)& 0.91(22) & 0.43(8) & 1.1(3) & 0.38(8) & 0.72 \\
      &0.80 & 0.425(5) & 0.78(14) & 0.40(5) & 3.6(6) & 0.33(5) & 0.82 \\%
      \cline{2-8}\multicolumn{3}{c}{}\\[-3mm]\cline{2-8}%
      &$\beta$ & $\kappa_c$ & $\beta_\chi$ & $\delta$ & $R$ & $S$ & $\chi^2$ \\
      \cline{2-8} $\cbcex$:
      &0.00 & 0.983(12) & 0.56(5)  & 3.1(3) & 1.3(2)  & 1.1(3)  & 0.84 \\
      &0.80 & 0.429(7)  & 0.56(10) & 3.0(5) & 4.4(12) & 2.4(13) & 0.58 \\
      \cline{2-8}
    \end{tabular}
    \caption{%
      Results of fits at $\beta=0.00$ and 0.80 using the equations
      of state. The upper table shows the results of the fit of $am_F$
      based on eq.~(\protect\ref{eq:eosmf}), the lower table those of
      $\cbcex$ based on eq.~(\protect\ref{eq:eoscbc}).}
    \label{tab:eossep}
  \end{center}
\end{table}

As a next step we performed a simultaneous fit with one common
$\kappa_c$ for $am_F$ and $\cbcex$ (table \ref{tab:eoskc}). A very
good fit to all the data was obtained.

Then we checked the scaling relations (\ref{eq:scal}).  Calculating
$\beta$ and $\delta$ with $\nu$ and $\tilde\nu$ gives $\beta_\chi =
0.54(20)$ and $\delta=3.8(15)$ for $\beta=0.00$ and
$\beta_\chi=0.39(25)$ and $\delta=5(3)$ for $\beta=0.80$.  The
agreement with the fit is quite good. Note that in (\ref{eq:scal}),
$d\tilde\nu=3\tilde\nu$ is close to 1 and hence the statistical errors
are increased.
\begin{table}
  \begin{center}
    \renewcommand{\arraystretch}{1.2}
    \begin{tabular}{|l|llllll|}
      \hline
      $\beta$ & $\kappa_c$ & $\nu$ & $\tilde\nu$ & $\beta_\chi$ &
      $\delta$ & $\chi^2$ \\
      \hline
      0.00 & 0.983(12) & 0.88(8) & 0.42(3) & 0.56(5) & 3.1(3) & 0.71 \\
      0.80 & 0.425(4) & 0.78(11) & 0.40(4)  & 0.47(5) & 3.4(3) & 0.71 \\
      \hline
    \end{tabular}
    \caption{%
      Results of fits of $am_F$ and $\cbcex$ at $\beta=0.00$ and
      0.80, using both equations of state with a common $\kappa_c$.
      }
    \label{tab:eoskc}
  \end{center}
\end{table}

\begin{figure}
  \begin{center}
    \fdiifig{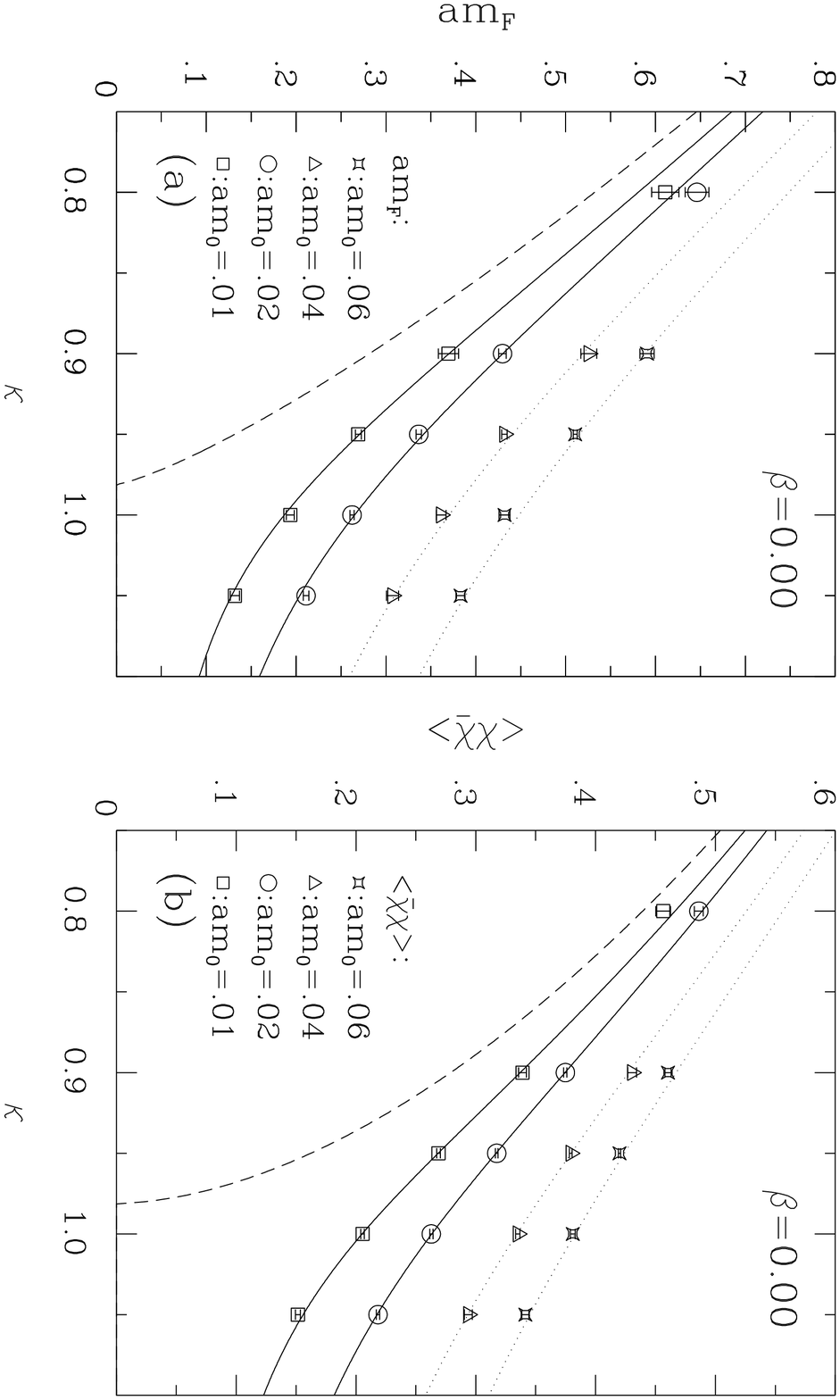}%
    \vspace{\capsep}%
    \caption[xxx]{%
      (a) Fermion mass and (b) chiral condensate for $\beta=0.00$. The data
      are our extrapolation into the infinite volume. The fit assumes the
      validity of the scaling relations and is described in the text.  The
      parameters are given in table \protect\ref{tab:scal}.  The dashed line
      shows the extrapolation into the chiral limit.}
    \label{fig:scal000}
    \vspace{5mm}
    \fdiifig{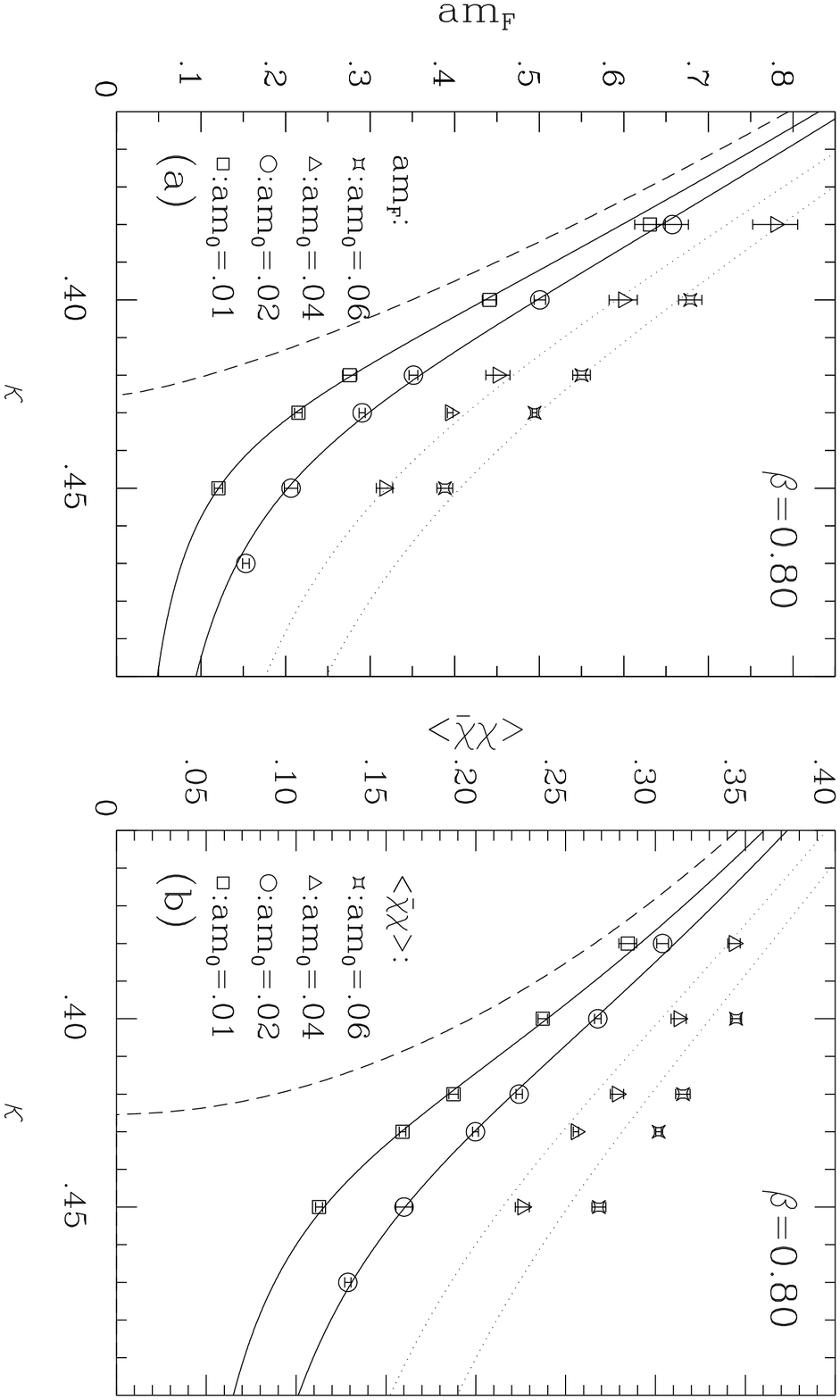}%
    \vspace{\capsep}%
    \caption[xxx]{%
      (a) Fermion mass and (b) chiral condensate for $\beta=0.80$.}
    \label{fig:scal080}
  \end{center}
\end{figure}

We also tried a third fit in which we assumed the validity of the
scaling relations (\ref{eq:scal}). The result is shown in
Figs.~\ref{fig:scal000} and \ref{fig:scal080} and summarized in
table~\ref{tab:scal}. As one can see, the quality of the fit is still
good and $\chi^2$ are reasonable.  The figures also show the
prediction of our fit for the fermion mass and chiral condensate at
$am_0=0.04$ and 0.06. Only small deviations are visible. We
therefore conclude that (\ref{eq:scal}) is consistent with our data.
\begin{table}
  \begin{center}
    \renewcommand{\arraystretch}{1.2}
    \begin{tabular}{|l|llll|ll|}
      \hline
      $\beta$ & $\kappa_c$ & $\nu$ & $\tilde\nu$ &
      $\chi^2$ & $\beta_\chi$ & $\delta$ \\
      \hline
      0.00 & 0.981(6) & 0.79(2) & 0.437(5) & 2.2 & 0.56(4) & 3.2(2) \\
      0.80 & 0.425(2) & 0.75(2) & 0.431(6) & 2.3 & 0.51(4) & 3.4(2)\\
      \hline
    \end{tabular}
    \caption{%
      Results of our fits using the equations of state at $\beta=0.00$ and
      0.80 with one $\kappa_c$ and the scaling relations
      (\protect\ref{eq:scal}) at $\beta=0.00$ and 0.80.} 
    \label{tab:scal}
  \end{center}
\end{table}

The values of the exponents $\nu$, $\tilde\nu$, and $\beta_\chi$ in
table \ref{tab:eossep} and table \ref{tab:eoskc} agree with those in
table \ref{tab:scal}. Thus all three fitting procedures gave
consistent results at each $\beta$. 

Furthermore, the exponents obtained at $\beta=0.00$ and $\beta=0.80$
agree within errors.  This is a strong signal that the chiral phase
transition is in one universality class at these $\beta$'s and
probably also for those in between.  The difference in the exponents
of the last fit, which is somewhat larger than the pure statistical
errors may for example be the result of small scaling deviations.

To estimate the uncertainty due to the choice of the extrapolation
formula, we repeated the above procedures using the extrapolation
(\ref{eq:1L}).  The results are given in table~\ref{tab:scal1L}.  The
$\chi^2$'s are even smaller and the exponents differ by a little more
than one standard deviation. Although the agreement for the two
$\beta$'s is less good, it is still compatible with universality if
one takes into account that the error bars only reflect the
statistical errors and not the uncertainty due to scaling deviations.
\begin{table}
  \begin{center}
    \renewcommand{\arraystretch}{1.2}
    \begin{tabular}{|l|llll|ll|}
      \hline
      $\beta$ & $\kappa_c$ & $\nu$ & $\tilde\nu$ &
      $\chi^2$ & $\beta_\chi$ & $\delta$ \\
      \hline
      0.00 & 0.968(9) & 0.76(3) & 0.426(8)  & 0.89 & 0.50(6) & 3.6(3)\\
      0.80 & 0.419(3) & 0.66(4) & 0.409(10) & 0.64 & 0.37(6) & 4.4(6)\\
      \hline
    \end{tabular}
    \caption[xxx]{%
      Results of fits, when we use eq.~(\ref{eq:1L}) for the
      extrapolation to infinite volume.  As in table \ref{tab:scal},
      the equations of state with common $\kappa_c$ and the scaling
      relations (\protect\ref{eq:scal}) are used.}
    \label{tab:scal1L}
  \end{center}
\end{table}

\subsection{The Finite Size Scaling of the Lee-Yang Zeros}

The Lee-Yang zeros were found to be purely imaginary at all values of
$\kappa$ in the strong coupling region. For small $\kappa$ they are
equally spaced, consistent with a strong first order transition in the
condensate as the bare mass $am_0$ goes through zero.

\begin{figure}
  \begin{center}
    \fdfig{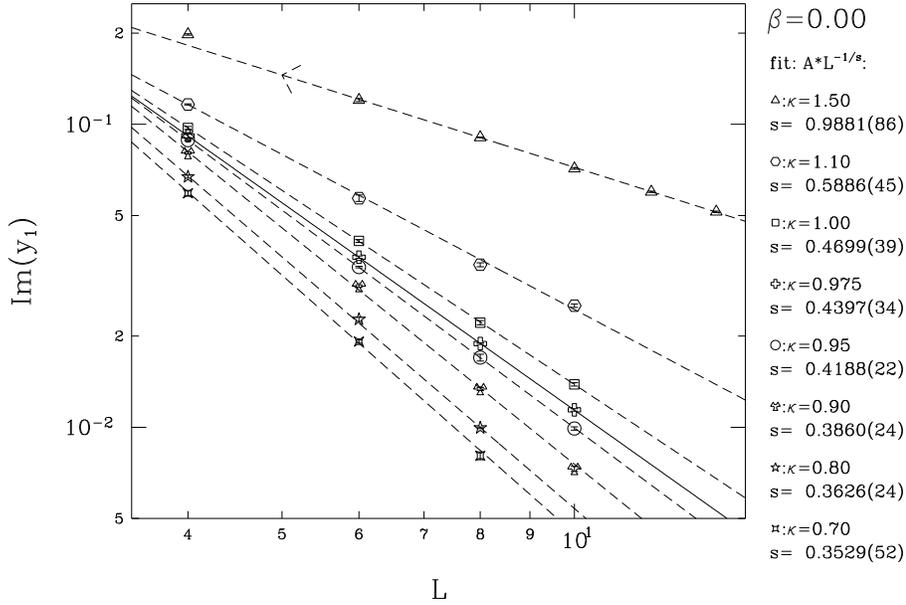}%
    \vspace{\capsep}%
    \caption[xxx]{%
      Imaginary part of zero $y_1$ as function of the lattice size for
      different $\kappa$ at $\beta=0.00$. The different straight lines
      should help to investigate the linearity and claim only for
      $\kappa=0.975 \approx \kappa_c$ (full line) to describe the data
      well.}
    \label{fig:ly000}
  \end{center}
\end{figure}

\begin{figure}
  \begin{center}
    \fdfig{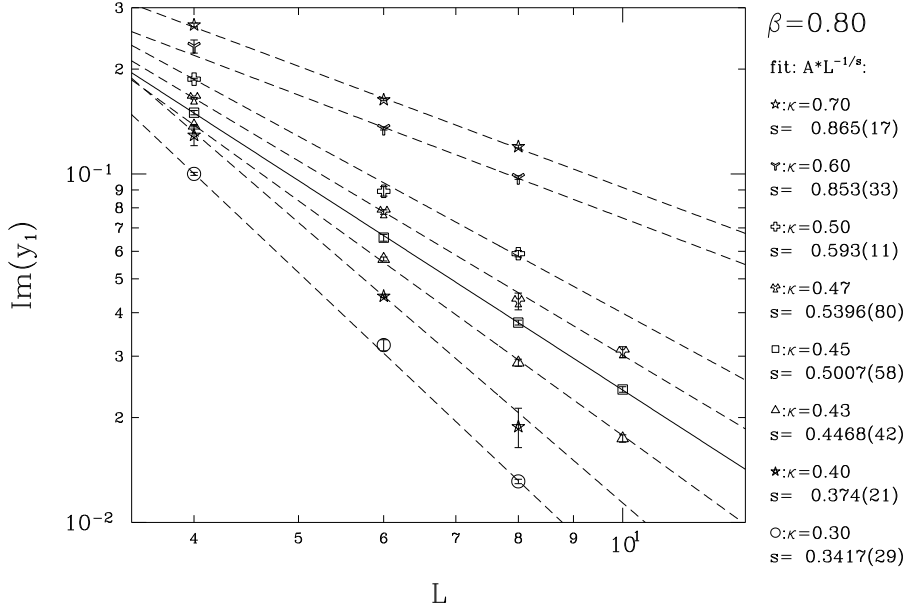}%
    \vspace{\capsep}%
    \caption[xxx]{%
      Imaginary part of zero $y_1$ as function of the lattice size for
      different $\kappa$ at $\beta=0.80$.}
    \label{fig:ly080}
  \end{center}
\end{figure}

Figs.~\ref{fig:ly000} and \ref{fig:ly080} show the finite size scaling
behaviour of the edge singularity at various $\kappa$ for $\beta=0.00$
and $0.80$, respectively. Our data confirm the expectations presented
in section \ref{sec:deflyz} and Fig.~\ref{fig:fsscal}. Close to the critical
point we see the expected crossover: for small lattices the exponent
is close to $\tilde\nu$ and shifts for increasing lattice size to the
exponent $1/3$ or $1$.


At small $\kappa$ the exponent $s$ is consistent with a first order
phase transition, $s \gapprox \frac{1}{3}$. At large $\kappa$ is $s
\lesssim 1$.  Close to the critical point, determined in the previous
section and given in table \ref{tab:scal}, the data scale linearly on
the log-log plot allowing determination of $\tilde\nu$.

At the $\kappa$ points closest to $\kappa_c$ we expect $s \simeq \tilde\nu$.
At $\beta=0.00$ we find at $\kappa = 0.975 \simeq \kappa_c = 0.981(6)$ the
exponent $s=0.440(4)$ in excellent agreement with $\tilde\nu=0.437(5)$, as
determined in the previous section.  For $\beta=0.80$ we did the simulations
at $\kappa = 0.43$ slightly larger than the $\kappa_c = 0.425(2)$ obtained
from the EOS.  Not unexpectedly we found $s=0.447(5)$, slightly larger than
$\tilde\nu=0.431(6)$ from the EOS.  This shows the great importance of the
knowledge of the critical point for a precision measurement of $\tilde\nu$.
Within these uncertainties the agreement is very good and again confirms the
independence of $\tilde\nu$ from $\beta$ and hence the universality.

\begin{figure}
  \begin{center}
    \fdfig{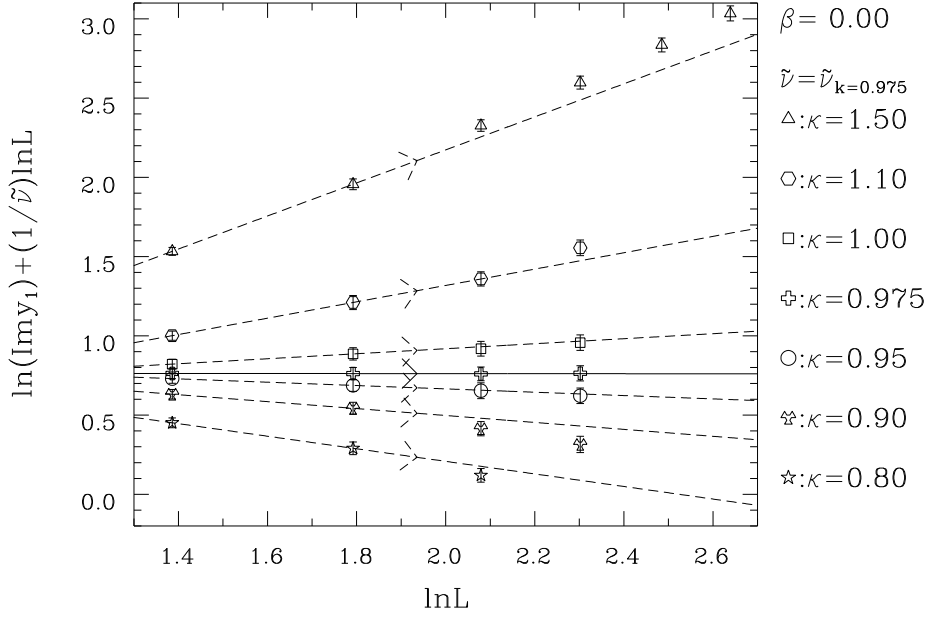}%
    \vspace{\capsep}%
    \caption[xxx]{%
      $\ln[Im(y_1)] + 1/\tilde\nu\ln L$ with $\tilde \nu = 0.440$ as a function
      of the lattice size for different $\kappa$'s at $\beta=0.00$. The dashed
      straight lines are linear extrapolations to the data points at the 
      lowest two values of $L$.}
    \label{linear0}
  \end{center}
\end{figure}

\begin{figure}
  \begin{center}
    \fdfig{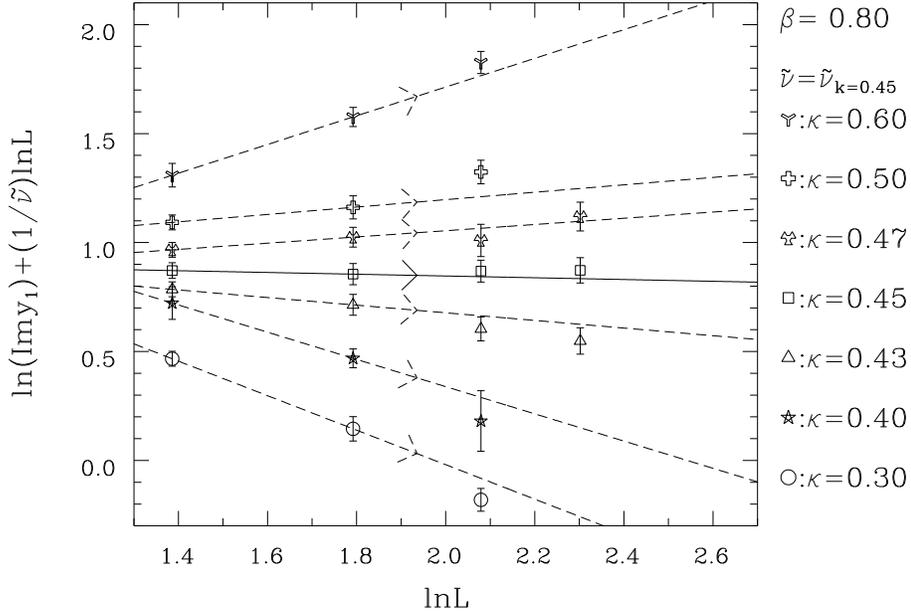}%
    \vspace{\capsep}%
    \caption[xxx]{%
      As in the previous figure but at $\beta=0.80$ with $\tilde \nu =0.501$.}
    \label{linear.8}
  \end{center}
\end{figure}

If one analyses these plots without the knowledge of the critical
point determined with the EOS, the critical point can also be
determined by looking for linearity of $\ln[Im(y_1)]$ as a function of
$\ln L$. For this purpose we show in Figs.~\ref{linear0} and
\ref{linear.8} the quantity $\ln[Im(y_1)] + 1/\tilde\nu\ln L$. The
addition of the second term makes the plots approximately horizontal
and so allows us to enhance the vertical scale making the error bars
clearer. In the figures we have used $\tilde \nu = 0.440$ and $0.501$,
respectively.  The dashed lines are a linear extrapolation of the data
points at the two lowest $L$-values. They provide a guide as to the
linearity of the data.


Figs.~\ref{linear0} and \ref{linear.8} suggest a larger $\kappa_c$ and
$s(\kappa_c)$ than those obtained from the EOS analysis.  For example,
at $\beta=0.80$, Fig.~\ref{linear.8} would suggest $\kappa = 0.45$ as
the point closest to $\kappa_c$, with $\tilde\nu \simeq s(0.45) =
0.501(6)$. However, the difference between the two methods of analysis
is about 10\% which is the same size as the statistical error.

All in all, this demonstrates the reliability of the methods we have
used.  Both (very different) methods agree rather well and their
combination is very useful.

\subsection{Universality at strong coupling}
Our data are a good indication that the chiral phase transition of the
\chupiii\ model is in the same universality class at $\beta=0.00$ and
$\beta=0.80$. Assuming this universality we combine the results for
exponents at both $\beta$ values and determine the exponents of this
chiral phase transition to be $\nu=0.75(10)$ and $\tilde\nu=0.43(2)$.
The errors take into account the uncertainties discussed above.  These
values of $\nu$ and $\tilde\nu$ correspond to $\beta_\chi=0.51(11)$
and $\delta=3.45(71)$. We note that the position of the critical point
at $\beta = 0$, as well as the results for $\delta$ and $\beta$ are
compatible\footnote{We thank S.J. Hands for pointing out to us this
  compatibility.} with those obtained for the $N_f=2$ case in
\cite{DeHa97} ($\beta_\chi=0.57(2)$ and $\delta=2.75(9)$). In that
work the same action (\ref{S-4-fermion}) has been simulated, though in
a somewhat different representation by means of auxiliary fields than
the $\beta=0$ limit of the \chupiii\ model.

It is very likely that the chiral phase transition is in the same
universality class for $\beta$ between 0 and the onset of the X region
around $\beta \simeq 1$. The universality might be expected at small
$\beta$ because of the convergence of the strong coupling expansion.
But our data are (to our knowledge) the first indication that this is
true for a large $\beta$ interval.

This result indicates that the \chupiii\ model is renormalizable in this
region of $\beta$. So it is a nontrivial example in three dimensions for the
shielded gauge-charge mechanism of fermion mass generation proposed in
\cite{FrJe95a}.

We note that, in the scaling investigation, the chiral condensate,
which is a pure fermionic operator, and the mass of the fermion $F$,
which is a combination of the fermion and the scalar field, have been
used. Both seem to scale in a way which can be well described by the
usual scaling relations.

The universality on the other hand also means that, with respect to
the three-dimensional Gross-Neveu model, nothing substantially new
happens at small and intermediate $\beta$ and no new physics arises on
scales much below the cutoff. The scalar field shields the fermion
$\chi$ giving rise to the fermion $F$ equivalent to the fermion of the
four-fermion theory. We find no indication that composite states
consisting only of fundamental scalars or gauge fields, which would
not fit into the Gross-Neveu model, scale at the chiral phase
transition.

The bosonic fields appear to be auxiliary at strong gauge coupling, as
they are in a rigorous sense \cite{LeShr87a} at $\beta = 0$.  As
indicated by the results in four dimensions \cite{FrJe98a,FrJe98b},
this may change as the gauge coupling gets weaker. We therefore
performed some studies at larger values of $\beta$. The results are
described in the next section.

\section{Explorative study of the weak coupling region}
\label{sec:weak}

\subsection{Condensate and fermion mass}
We should point out again that the compact three-dimensional QED with
fermions is not fully understood at large $\beta$. It is not clear, at
large $\beta$, if there is chiral symmetry breaking and confinement
via a linear potential. This uncertainty extends also to our model
with $\kappa$ small. A clarification of these difficult questions
would require a substantial effort far beyond the scope of this paper.
Thus our aim is to perform an explorative study only and to get some
insight into this as yet unexplored region. Also we want to see how
far the methods applied successfully at strong coupling can be of use
also at weaker coupling. The physical interpretation of our results
will leave room for several scenarios.

\begin{figure}
  \begin{center}
    \leavevmode
    \fdiifig{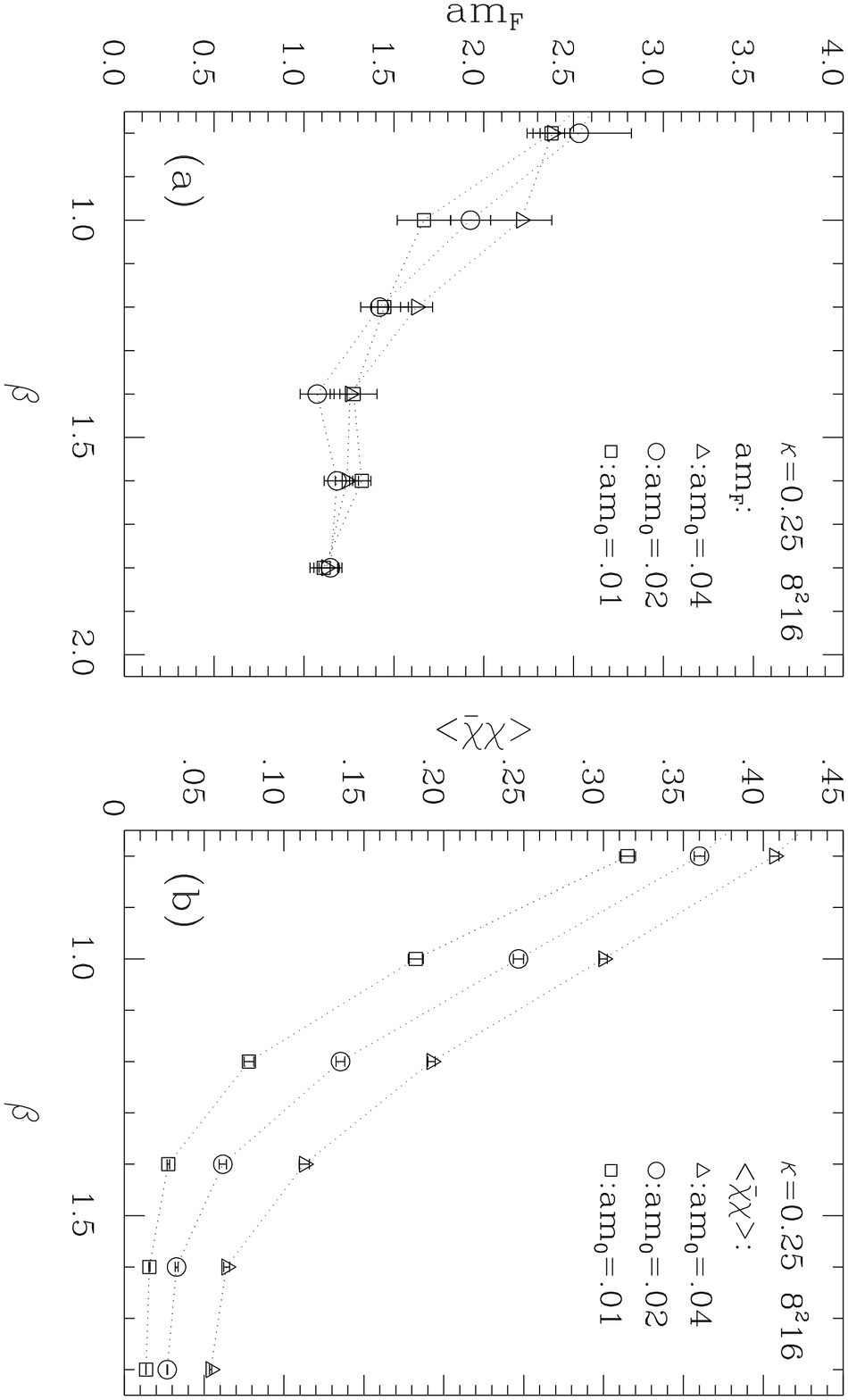}%
    \vspace{\capsep}%
    \caption[xxx]{%
      (a) fermion mass and (b) chiral condensate for different small $am_0$ as
      function of $\beta$ at $\kappa=0.25$ on the $8^216$ lattice.}
    \label{fig:fermcbck015}
  \end{center}
\end{figure}

Fig.~\ref{fig:fermcbck015} shows the fermion mass and condensate at
$\kappa=0.25$ as a function of $\beta$, at three values of the bare
fermion mass.
The neutral fermion mass decreases for increaing $\beta$ but then stabilizes
with $am_F>1$. So it is clearly nonzero at all $\beta$ and again only weakly
dependent on the bare fermion mass. The condensate is large at small $\beta$
(the Nambu phase) but rapidly decreases with $\beta$ and becomes very small
(zero?) in the chiral limit for $\beta > \beta_X \simeq 1.3$. Thus here a
new, weakly coupled region is encountered.

\begin{figure}
  \begin{center}
    \fdiifig{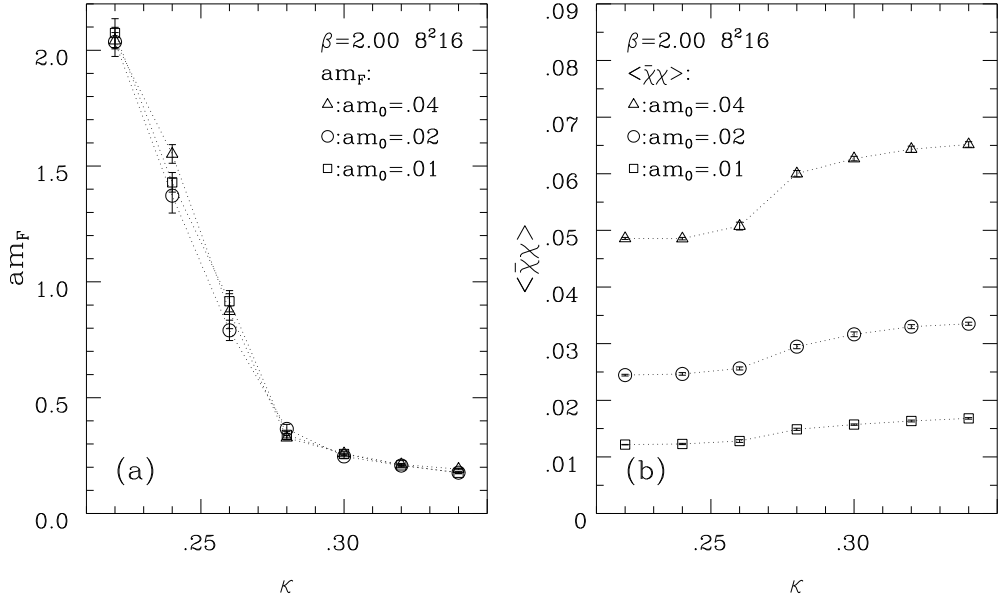}%
    \vspace{\capsep}%
    \caption[xxx]{%
      (a) fermion mass and (b) chiral condensate for different small
      $am_0$ as function of $\kappa$ at $\beta=2.00$ on the
      $8^216$ lattice.}
    \label{fig:fermcbcb200}
  \end{center}
\end{figure}

In order to see how this region is related to the Higgs phase at large
$\kappa$, the fermion mass\footnote{We shold remark that for large $\beta$
  the agreement of the different fits for the fermion mass is not as good as
  that at small $\beta$.  The qualitative behaviour is not influenced by this.
  This same feature was also observed in the four dimensional model for large
  $\beta$ but is not understood up to now.} and condensate are shown in
Fig.~\ref{fig:fermcbcb200} at $\beta=2.0$ and three values of the bare fermion
mass. For nonvanishing bare mass, where the simulations have been performed,
the mass of the neutral fermion is large for $\kappa < \kappa_X \simeq 0.27$
whereas it is small for larger $\kappa$.  It is only very weakly dependent on
the bare fermion mass and therefore we expect this behaviour to persist in the
chiral limit.  For $\kappa>\kappa_X$ its small nonzero value probably vanishes
in the infinite lattice size limit.

The condensate, as expected, does depend strongly on the bare mass
$am_0$ but does show a crossover behaviour at $\kappa = \kappa_X$ with
$\cbcex(am_0)_{\kappa<\kappa_X} < \cbcex(am_0)_{\kappa>\kappa_X}$.
However, at large $\kappa$ we believe that we are in the Higgs phase
where the condensate is zero in the chiral limit. It is therefore
conceivable that, in this limit, it is zero at $\beta=2.0$ for all
$\kappa$.  It is very surprising, however, that, at fixed bare fermion
mass and lattice size, the condensate tends to slightly increase with
increasing $\kappa$, quite in contrast from its behaviour in the
strong coupling region. Of course, this can change in the infinite
volume and chiral limit.

At the coupling $\kappa = \kappa_X$, where the fermion mass shows a
possible crossover behaviour, there is also a peak in the
susceptibility of the link energy which increases with increasing
lattice volume. However, a careful finite size scaling analysis would
be needed to determine if it indicates a phase transition or a
crossover.

\begin{figure}
  \begin{center}
    \fdiifig{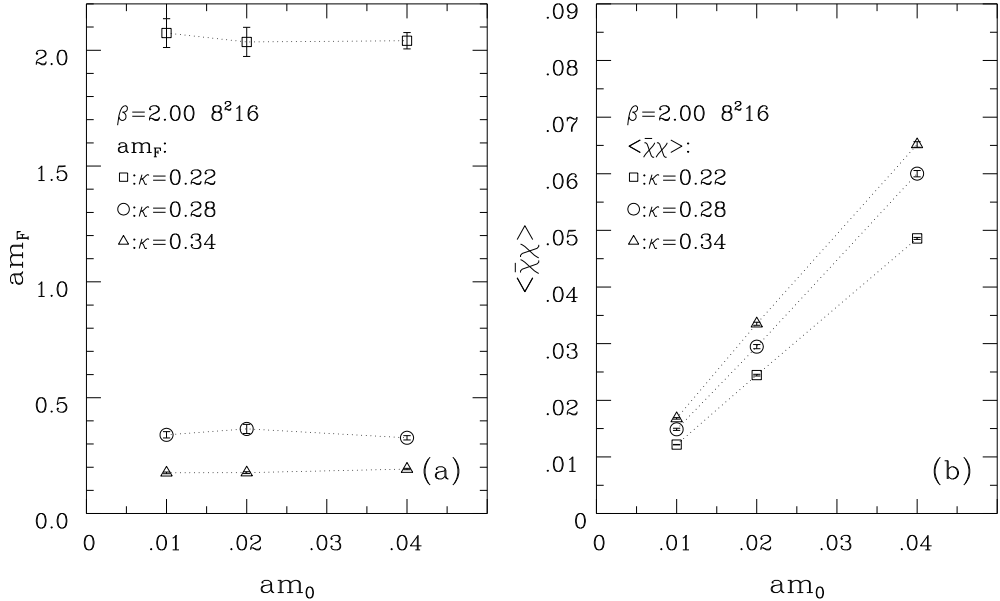}%
    \vspace{\capsep}%
    \caption[xxx]{%
      (a) Fermion mass and (b) chiral condensate for different $\kappa$ as
      function of $am_0$ at $\beta=2.00$ on the $8^216$ lattice.}
    \label{fig:m_fermcbc}
  \end{center}
\end{figure}

Fig.~\ref{fig:m_fermcbc}a confirms the weak dependence of the fermion
mass $am_F$ on the bare mass. At $\kappa = 0.22$ below $\kappa_X$, the
fermion mass is large and stays clearly nonzero at $am_0$.  Above
$\kappa_X$, at $\kappa = 0.34$, the fermion mass is too small for the
lattice size used and might vanish in the infinite volume limit.

Fig.~\ref{fig:m_fermcbc}b shows that, at $\beta=2.0$ and $\kappa$ just
below or above $\kappa_X$, the condensate extrapolates linearly in
$am_0$ to a very small value or zero. As we shall see below, this is
due to the Lee-Yang edge singularity in this region being relatively
distant from the real axis.

A naive extrapolation to the chiral limit would thus classify the region at
small $\kappa$ ($\kappa < \kappa_X$) and large $\beta$ ($\beta > \beta_X$)
as a phase with zero chiral condensate and nonvanishing fermion mass. But the
condensate could also remain very small but nonvanishing. Because of this
uncertainty we label this region X. Its boundaries $\beta_X$ and $\kappa_X$
may slightly depend on $\kappa$ and $\beta$, respectively.

It would be surprising if in the region X the fermion mass was different from
zero with unbroken chiral symmetry. There are essentially two scenarios
avoiding such a paradox.

1) Chiral symmetry breaking persists at small $\kappa$ also for $\beta
> \beta_X $. $\chi$ is light, because the chiral condensate is very
small, though nonzero. $F$ is a bound state of $\phi$ and $\chi$. The
binding might be quite loose, presumably by a weak linear confining
potential, which one expects in pure U(1) in 3d
\cite{Po77,BaMy77,GoMa82}. $F$ is heavy essentially because $\phi$ is
heavy. The transition at $\beta = \beta_X$ is probably a cross-over,
but a genuine phase transition is not excluded.

In this scenario the region X must be separated by a chiral phase
transition at $\kappa_X(\beta)$ from the Higgs phase. As our data
around $\kappa _X$ do not indicate any metastability, it would be a
higher order transition and a continuum limit should be possible. Thus
an interesting continuum limit with massive unconfined fermion might
exist. 

2) Chiral symmetry is restored at $\beta = \beta_X$ and the chiral
condensate thus vanishes identically and $\chi$ is massless in X. The
$F$ channel gets contribution from the two-particle state $\phi$ and
$\chi$. This contribution appears as a massive state because $\phi$ is
heavy. This state presumably cannot be a bound state in the chiral
limit because of the old argument of Banks and Casher \cite{BaCa80}:
fermion on a closed orbit must be able to flip its helicity, i.e.
existence of the bound state implies chiral symmetry breaking. We
cannot distinguish between a bound state $F$ and a two-particle state
$\phi$ + $\chi$ looking at the $F$ channel only (as we did). X could
be connected to the Higgs phase, where we expect the same spectrum.

Which of these scenarios is true might be investigated in the limit
case $\kappa = 0$, i.e. in the three-dimensional compact QED. The
results in the noncompact case \cite{ApNa88,DaKo89,DaKo90,HaKo90}
might be applicable at weak coupling also to the compact one. As the
number of fermions in our case is below the critical number $\simeq
3.5$ of fermions in the noncompact model, the more interesting
scenario 1) seems to be preferred.

\subsection{The Lee-Yang zeros at Weak Coupling}
\begin{figure}
  \begin{center}
    \fdiifig{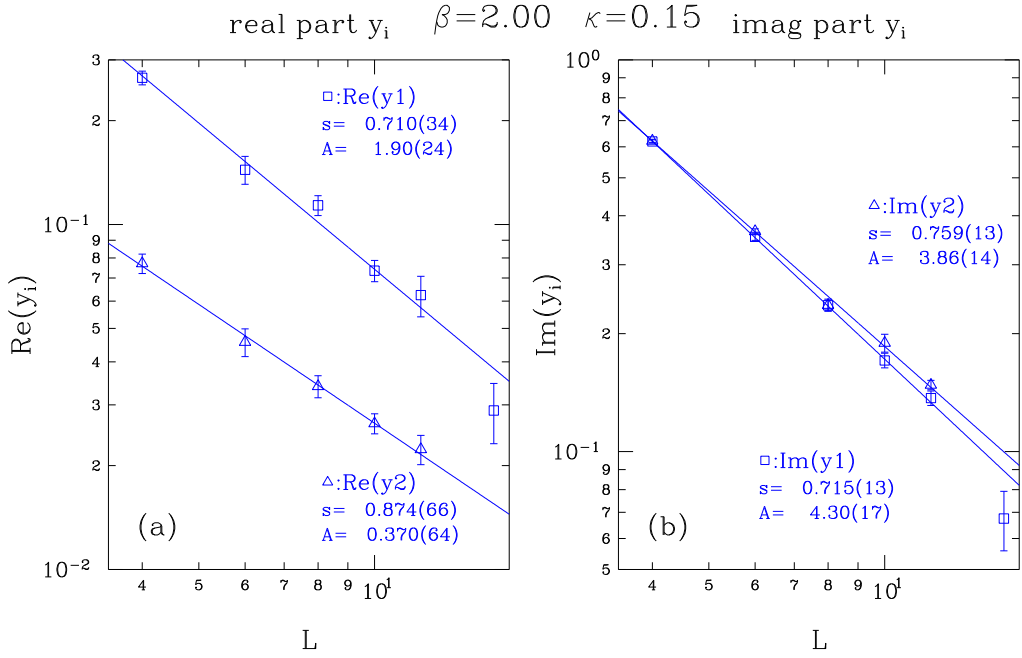}%
    \vspace{\capsep}%
    \caption[xxx]{%
      (a) Real and (b) imaginary part of the first two Lee Yang zeros
      (sorted by their positive imaginary part) for $\beta=2.00$,
      $\kappa=0.15$ as function of lattice size.}
    \label{fig:reimy200}
  \end{center}
\end{figure}

In an attempt to clarify the situation at weak coupling we have also
investigated the Lee-Yang zeros in the region X. The edge
singularity\footnote{The zeros must appear in conjugate pairs. We
  define the edge singularity in this region to be the zero with
  smallest positive imaginary part and count each pair only once.}
$y_1$ has a nonzero real part in this region.  The finite size scaling
of the lowest zeros is shown in Fig.~\ref{fig:reimy200} for
$\beta=2.00$, $\kappa=0.15$, a point in the middle of the region X.
The real part of the low lying zeros is clearly nonvanishing.  Then
the first two zeros have within the numerical precision identical
imaginary part but their real parts differ by a factor of about 3.5.

The first two zeros have imaginary parts so close as to be
indistinguishable within statistical error. We have assumed continuity
in the behaviour of their real parts as a function of lattice size
when plotting Fig.~\ref{fig:reimy200}.

These imaginary parts scale linearly in the log-log plot with an
exponent $s=0.73(4)$ where the error is given by the difference
between the two zeros.  Their real part has somewhat larger errors but
scales within the numerical precision with the same exponent. This
pattern for the edge singularity is found throughout the region X.
Note that this behaviour is only observed in the edge singularity.

\begin{figure}
  \begin{center}
    \fdfig{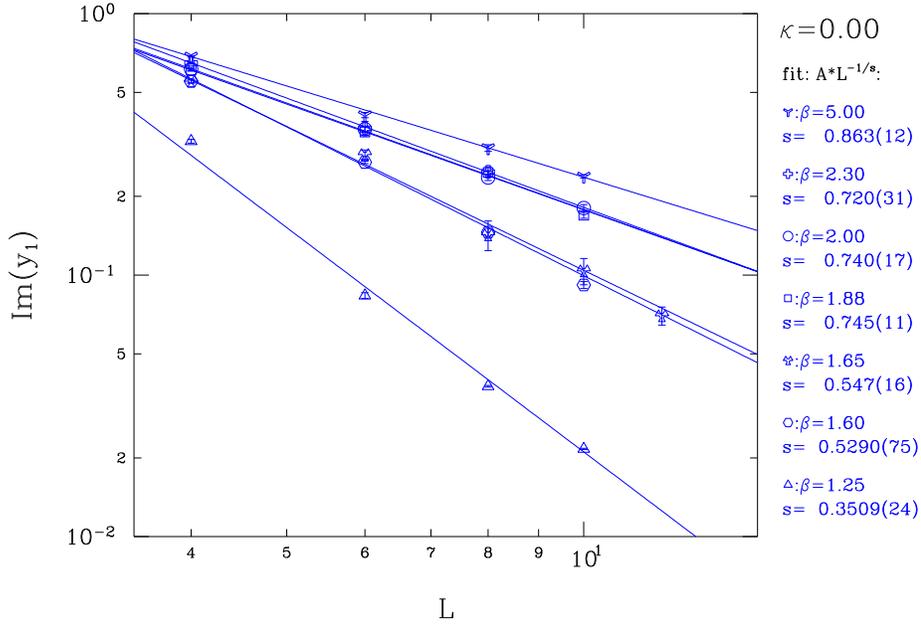}%
    \vspace{\capsep}%
    \caption[xxx]{%
      Imaginary part of zero $y_1$ as function of the lattice size for
      different $\beta$ at $\kappa=0$.}
    \label{fig:lyk000}
  \end{center}
\end{figure}%
Fig.~\ref{fig:lyk000} shows the behaviour of the imaginary part of the
edge as a function of lattice size at $\kappa=0$ for various $\beta$.
There is a crossover between $\beta=1.25$ and $\beta=1.88$, i.\,e.
from the Nambu phase, where the imaginary part of the edge zero is
small and the transition first order, to a region where the imaginary
part is large (with nonzero real part) again consistent with a
vanishing condensate. No scaling deviations can be observed for
$\beta\geq 1.88$. In the region X the critical exponent has a very
weak dependence on $\beta$ and increases only very slowly on further
increase of $\kappa$.

If the scaling in the region X is different from that in the other
regions, the most naive expectation would be that the exponent $s$ is
universal. This is compatible with our data at $\beta\approx 2$ but
not at $\beta=5$.  Further simulations on larger lattice are necessary
to confirm this difference.

It is to our knowledge the first model in which scaling of the real
part of the edge singularity to zero has been observed. We do not
understand the implications of this behaviour. It may well be a key
point in understanding the critical nature of region X.

Summarizing, the region X can be distinguished from the Nambu and
Higgs phase by the edge singularity having a real part (on a finite
lattice) and a scaling which cannot be described by either an exponent
$s=1/3$ or $s=1$. If the exponent $s$ is different from those in Nambu
and Higgs phases then the region X is presumably a new phase. We have
been unable to determine if the chiral symmetry is broken or not in
this region.

\section{Conclusions}
We have presented an extensive analysis of the phase structure of the
three-dimensional fermion-gauge-scalar model.  The analysis has been
made possible by the application of two different methods: $1$)fits to
an equation of state of the chiral condensate and the mass of the
physical neutral fermion and $2$)finite size scaling investigations of
the Lee-Yang zeros of the partition function in the complex fermion
mass plane.

Our investigations showed that there are three regions in the
$\beta$-$\kappa$ plane with possibly different critical behaviour in
the chiral limit:

a) The region at small $\kappa$ and strong gauge coupling, where the
chiral symmetry is broken and the neutral physical fermion is massive,
called the Nambu phase.

b) The region at large $\kappa$, where the chiral symmetry is restored
and the physical fermion is massless, called the Higgs phase.

c) A third region at weak coupling and small $\kappa$, where the
chiral condensate is zero within our numerical accuracy but the
neutral fermion mass is large, called the X region.  This region can
analytically be connected with either the Nambu or Higgs phase but it
may well be a separate phase. If chiral symmetry is not broken in this
region, then the mass observed in the fermion channel is presumably
the energy of a two-particle state. Otherwise this region might be an
interesting example of dynamical mass generation of unconfined
fermions. If the continuum limit is taken at the Higgs phase
transition, the gauge fields should play an important dynamical role
and the model would not fall into the universality class of the
three-dimensional Gross-Neveu model. A further investigation of this
possibility is highly desirable.

At strong gauge coupling, the chiral phase transition can be clearly
localized, and there are strong indications that it is in one
universality class for all $\beta < 1$: that of the three-dimensional
Gross-Neveu model, which is known to be non-pertubatively
renormalizable. This demonstrates that the three-dimensional lattice
\chupiii\ model is a nonperturbatively renormalizable quantum field
theory and the shielded gauge-charge mechanism of fermion mass
generation \cite{FrJe95a} works in three dimensions.

\subsection*{Acknowledgements}
We thank M.~G\"ockeler, S.~J. Hands, and K.-I.~Kondo for discussions, and
J. Paul for valuable contributions in the early stage of this work.
The computations have been performed on the Fujitsu computers at RWTH
Aachen, and on the CRAY-YMP and T90 of HLRZ J\"ulich.  E.F., W.F., and
J.J. acknowledge the hospitality of HLRZ.

The work was supported in part by DFG.  I.B. was supported in part by
the TMR-network "Finite temperature phase transitions in particle
physics", EU-contract ERBFMRX-CT97-0122.

\bibliographystyle{wunsnot}   


\end{document}